# Photo-molecular high temperature superconductivity

M. Buzzi[1,*], D. Nicoletti[1], M. Fechner[1], N. Tancogne-Dejean[1], M. A. Sentef[1],
A. Georges[2,3], M. Dressel[4], A. Henderson[5], T. Siegrist[5], J. A. Schlueter[5,6], K. Miyagawa[7],
K. Kanoda[7], M.-S. Nam[8], A. Ardavan[8], J. Coulthard[8], J. Tindall[8], F. Schlawin[8], D. Jaksch[8],
A. Cavalleri[1,8,*]

[1] Max Planck Institute for the Structure and Dynamics of Matter, 22761 Hamburg, Germany
[2] Center for Computational Quantum Physics (CCQ), The Flatiron Institute, New York, NY, USA
[3] Collège de France, 11 place Marcelin Berthelot, 75005 Paris, France
[4] 1. Physikalisches Institut, Universität Stuttgart, 70569 Stuttgart, Germany
[5] National High Magnetic Field Laboratory, 1800 E Paul Dirac Drive, Tallahassee, FL 32310, USA
[6] Division of Material Research, National Science Foundation, Alexandria, VA 22314, USA
[7] Department of Applied Physics, University of Tokyo, 7-3-1 Hongo, Bunkyo-ku, Tokyo 113-8656, Japan
[8] Department of Physics, Clarendon Laboratory, University of Oxford, Oxford OX1 3PU, United Kingdom
* e-mail: michele.buzzi@mpsd.mpg.de, andrea.cavalleri@mpsd.mpg.de

**Superconductivity in organic conductors is often tuned by the application of chemical or external pressure. With this type of tuning, orbital overlaps and electronic bandwidths are manipulated, whilst the properties of the molecular building blocks remain virtually unperturbed. Here, we show that the excitation of local molecular vibrations in the charge-transfer salt κ-(BEDT-TTF)$_2$Cu[N(CN)$_2$]Br induces a colossal increase in carrier mobility and the opening of a superconducting-like optical gap. Both features track the density of quasi-particles of the equilibrium metal, and can be achieved up to a characteristic coherence temperature $T^* \simeq 50$ K, far higher than the equilibrium transition temperature $T_C$ = 12.5 K. Notably, the large optical gap achieved by photo-excitation is not observed in the equilibrium superconductor, pointing to a light induced state that is different from that obtained by cooling. First-principle calculations and model Hamiltonian dynamics predict a transient state with long-range pairing correlations, providing a possible physical scenario for photo-molecular superconductivity.**

Organic conductors based on the BEDT-TTF (bisethylenedithio-tetrathiafulvalene) molecules display low dimensional electronic structures and unconventional high-$T_C$ superconductivity (*1-3*). In the crystals with κ-type arrangement, the (BEDT-TTF)$^{+0.5}$ molecules (henceforth abbreviated as ET) form dimers, which are organized in planes on a triangular lattice (*4*). Due to dimerization, the quarter-filled band originating from the overlap of the molecular orbitals splits into pairs of half-filled bands. The ET planes are separated by layers of monovalent anions, which act as charge reservoirs (Fig. 1(a)). The compound with the highest superconducting transition temperature is κ-(ET)$_2$Cu[N(CN)$_2$]Br (henceforth abbreviated as κ-Br), for which $T_C$ = 12.5 K.

The low-energy physics of the κ-phase compounds is captured by a Hubbard model on a triangular lattice with weak and strong bonds (Fig 1(a), lower right panel) (*5-7*). The ground state can be tuned either by applying hydrostatic pressure or by chemical substitution with different anions (*8*). In both cases, the spacing between the dimers changes (Fig 1(a)), directly acting on the hopping integrals ($t$, $t'$), leaving the on-site wavefunction and electronic correlations essentially unperturbed. Here, we explore the possibility of dynamically manipulating the Hamiltonian parameters and the many-body wavefunction by resonant excitation of ET molecular vibrations (Fig. 1(b)).

Equilibrium optical spectra were measured in the normal state of κ-Br single crystals using a Fourier transform spectrometer for temperatures between 15 K and 300 K (see Fig. 2 and Supplementary Material), in good agreement with those reported previously in the literature (*9, 10*). In the equilibrium normal state ($T > T_C$), κ-Br is a quasi-two-dimensional Fermi liquid, characterized by a narrow Drude peak in the in-plane optical conductivity, observed up to $T^* \simeq 50$ K (*11, 12*). For temperatures $T > T^*$ the quasi-particle response vanishes and the material exhibits the behavior of a so-called "bad metal" (see Fig. 2(b)). A large gap and a temperature-independent insulating optical conductivity are found along the perpendicular (interlayer) direction (*13*) (Fig. 2(a)). Several narrow peaks are also evident in the mid-infrared spectral region, corresponding to infrared-active vibrational modes of the ET molecules (*14-16*).

Femtosecond mid-infrared pump pulses were polarized along the out-of-plane crystallographic axis and tuned to the spectral region where the molecular modes are found ($\nu_{pump} \simeq 900 - 2000$ cm$^{-1}$, $\lambda_{pump} \simeq 5 - 11$ µm). The peak pump electric field was varied between ~500 kV/cm and ~4 MV/cm. The changes in low-frequency reflectivity and complex optical conductivity induced by mid-infrared excitation were measured for frequencies between ~0.8 and 7 THz with probe pulses polarized along the conducting layers. These were detected by electro-optical sampling after reflection from the sample at different pump-probe time delays.

Figure 3 shows the most suggestive results discussed in this manuscript. Therein, optical properties ($R(\omega)$, $\sigma_1(\omega) + i\sigma_2(\omega)$) are reported for four representative base temperatures between $T = 15$ K and $T = 70$ K, measured before (red) and 1 ps after photo-excitation (blue). Here, the pump pulses were tuned close to resonance with molecular vibrations corresponding to distortions of the C=C bonds on the ET dimers (see sketch in Fig. 2). Importantly, these pump pulses, polarized along the insulating direction, penetrated here deeper than the THz probe field (polarized along the metallic planes, see also Supplementary Material). Hence, from the "raw" changes in electric field reflectance, $r_1(\omega) + ir_2(\omega)$, one could directly extract the complex optical conductivity ($\sigma_1(\omega) + i\sigma_2(\omega)$) without the need to account for an inhomogeneously excited medium (*17-22*).

The transient optical properties measured at base temperature $T = 15$ K, that is immediately above $T_C = 12.5$ K, exhibited the largest changes. The reflectivity saturated to $R = 1$ over a broad frequency range, reducing only above ~ 120 cm$^{-1}$.

Correspondingly, a gap opened in $\sigma_1(\omega)$ and a $\sim 1/\omega$ divergence developed in $\sigma_2(\omega)$. A qualitatively similar response was recorded for $T = 30$ K and $T = 50$ K, although with smaller gap and $\sigma_2(\omega)$ divergence. For these three temperatures, the metallic optical spectra measured at negative time delays were fitted with a Drude-Lorentz model (red lines in Fig. 3) for normal conductors, whereas all the transient optical properties were fitted with an extension of the Mattis-Bardeen model for superconductors of variable purity (*23, 24*) (blue lines in Fig. 3, see also Supplementary Material).

A qualitatively different response was measured at $T = 70$ K, for which $\sigma_1(\omega)$ increased rather than decreasing to zero, and $\sigma_2(\omega)$ remained characteristic of a metal without diverging toward low frequencies. These 70 K spectra were then fitted with the same Drude-Lorentz model used to reproduce the equilibrium response.

For each of the three measurements at $T \leq 50$ K reported in Fig. 3, we extracted the optical gap, $2\Delta$, from the Mattis-Bardeen fits. The temperature dependent gap size could be fitted by the function $\Delta(T) = \Delta(0) \tanh\left[\beta\sqrt{\frac{T'-T}{T}}\right]$, where $2\Delta(0) = 23$ meV was the zero-temperature non-equilibrium gap, $\beta = 1.74$, and $T' \simeq 52$ K provided an effective "critical" temperature for the non-equilibrium state (see Fig. 4(a)). Note that $T' \sim T^* \simeq 50$ K, which coincides with the temperature at which a coherent Drude peak was observed in the equilibrium metallic state of Fig. 2.

The connection between the superconducting-like optical properties measured in the transient state and the Fermi-liquid behavior of the equilibrium normal state (*12, 25*) is further underscored by the analysis reported in Fig. 4(b). Therein, the divergence in transient $\sigma_2(\omega)$ was used to estimate the light-induced "superfluid density", determined as $N_{eff}^{Trans} = \frac{mV_{Cell}}{e^2} \lim_{\omega \to 0}[\omega \sigma_2^{Trans}(\omega)]$. This quantity was compared to the temperature dependent density of equilibrium quasi-particles, $N_{eff}^{Equil} = \frac{mV_{Cell}}{4\pi e^2}(\omega_P^{Equil})^2$.

Here, $m$ is the bare electron mass, $V_{Cell}$ the unit cell volume, $e$ the electron charge, and $\omega_P^{Equil}$ the equilibrium carrier plasma frequency (extracted from Drude-Lorentz fits). Even though the absolute values on the two vertical axes of Fig. 4(b) are different, the temperature dependences are similar, with the same onset at $T^* \simeq 50$ K.

The time dependence of these features is visualized in Figure 5, where three different quantities are displayed. Fig. 5(a) reports the transient integrated spectral weight $\int_{\omega=40 \text{ cm}^{-1}}^{110 \text{ cm}^{-1}} \sigma_1(\omega) d\omega$, which in a superconductor exhibits a depletion by an amount proportional to the density of condensed pairs. Fig. 5(b) shows the transient superfluid density, $N_{eff}^{Trans}$, extracted from the low-frequency imaginary conductivity, as in Fig. 4(b). Fig. 5(c) reports $\sigma_0 = \lim_{\omega \to 0} \sigma_1(\omega)$, that is the extrapolated "zero-frequency" conductivity from Drude-Lorentz fits to the transient optical properties (see Supplementary Material). All three quantities changed promptly, immediately after photoexcitation and relaxed within a few picoseconds, a time scale that we tentatively associate here to the lifetime of the resonantly driven ET vibration. Importantly, $\sigma_0$ diverged to values larger than $10^5 \Omega^{-1} \text{cm}^{-1}$ (corresponding to mean free paths of at least ~100 unit cells), limited here by the 0.8 THz low-frequency cutoff of our measurement, which in turn was set by the relaxation time of the state (~ 2 ps).

In Figure 6, we report the transient complex optical conductivity after photo-excitation at different driving frequencies, all measured by keeping the pump fluence fixed to ~2 mJ/cm². A superconducting-like response could only be observed for excitation at 8 μm and 6.8 μm (see also Fig. 1(a)), while detuned driving to both red ($\lambda_{pump} = 11$ μm, resonant to a weaker ET vibration) and blue ($\lambda_{pump} = 5$ μm) sides resulted in a slight increase in $\sigma_2(\omega)$ and negligible changes in $\sigma_1(\omega)$. The $\sigma_0$ values extracted from these data are displayed in the right panel of Fig. 6, along with the equilibrium out-of-plane imaginary dielectric function. This plot confirms that superconducting like optical properties in the transient state can only be observed when pumping in the ~6.8 − 8 μm range, close to resonance with C=C modes of the ET molecule.

In searching for a microscopic mechanism for the observed photo-molecular response, we note that in a previous work on the one-dimensional Mott insulator ET-$F_2$TCNQ, featuring the same ET molecular building block as κ-Br, selective driving of an IR-active vibration was shown to provide a dynamical modulation of the on-site Hubbard-$U$ interaction, achieved by quadratic electron-phonon coupling (*26, 27*).

In the κ-Br system studied here, ab-initio calculations in the frozen-phonon approximation yielded an estimate for the on-dimer Hubbard $U$, hopping matrix elements $t$ (strong bonds), and $t'$ (weak bonds) as a function of normal mode distortion (see Supplementary Material). Excitation of the molecular C=C stretching modes (i.e. the $\nu_{27}$ vibration at 6.8 μm) leads to a significant modulation of $U$ and $t$, whereas $t'$ remains mostly unaffected (Fig. 7(a) and Supplementary Material). By contrast, excitation of molecular modes that involve mainly motions in the terminal ethylene groups of the ET molecule (*i.e.*, at 9 and 11 μm) does not produce any sizeable change in the effective electronic parameters (Fig. 7(b)).

Starting from these estimates, we performed simulations of a driven Fermi-Hubbard model on a triangular ladder system (see Supplementary Material for the Hamiltonian and specific geometry of the system). In the driven state, correlations between pairs of doublons residing on different sites were quantified by the doublon correlation function $\langle d_i^\dagger d_j \rangle = \langle c_{i\uparrow}^\dagger c_{i\downarrow}^\dagger c_{j\downarrow} c_{j\uparrow} \rangle$ (here, $c_{i\sigma}^\dagger$ and $c_{i\sigma}$ are single-particle creation and annihilation operators, respectively). In the simulations, the system was initialized in its half-filled zero-temperature ground state, yielding doublon correlations which decay exponentially with distance (blue dashed line in Fig. 7(a)). When the driving was turned on, long-range, uniform on-site doublons were formed with long correlation distances. These correlations stabilised and persisted for timescales compatible with those explored in the experiment (blue solid line, Fig. 7(a), see also Supplementary Material). When considering these theoretical predictions, a driving-induced emergence of long-range order in the system is hypothesized, where the external driving establishes both a large density of on-site doublons and phase coherence over macroscopic distances. The predictions reported in Fig. 7(b) for excitation of a terminal ethylene mode, for which no enhancement in coherent on-site doublons is expected, are in agreement with the experimental results of Fig. 6, which show that a coherent state is induced only for excitation of the C=C stretching mode.

The scenario proposed by the model sketched in Fig. 7 is notable as it implies the formation of a qualitatively new state of the system, with double occupancies that are created through the $U/t$ modulation. This mechanism, which may explain the appearance of a large gap not present in the low temperature equilibrium superconductor, bears some resemblance with the proposed photo-induced $\eta-$pair superconductor (*28-30*), in which repulsive pairs are activated by the drive.

Other mechanisms, different from that discussed here may also be at play (*31-33*), including the stabilization of pre-existing Cooper pairs in the normal state (*34*), which have been revealed by superconducting Nernst effect measurements (*35*). Finally, the possible appearance of a negative-$U$ superconductivity through electronic squeezing in a molecular excited state should also be considered, especially in the context of a nearly frustrated electronic structure like the present one (*36, 37*).

# Acknowledgments

The research leading to these results received funding from the European Research Council under the European Union's Seventh Framework Programme (FP7/2007-2013)/ERC Grant Agreement No. 319286 (QMAC). We acknowledge support from the Deutsche Forschungsgemeinschaft (DFG) via the Cluster of Excellence 'The Hamburg Centre for Ultrafast Imaging' (EXC 1074 – project ID 194651731), the priority program SFB925, and the Emmy Noether program (SE 2558/2-1). J. A. Schlueter acknowledges support from the Independent Research/Development program while serving at the National Science Foundation. K. Miyagawa and K. Kanoda acknowledge support from the Japan Society for the Promotion of Science Grant No. 18H05225. T. Siegrist and A. Henderson acknowledge funding from the NSF under grant DMR-1534818. The NHMFL is supported by the NSF under grant DMR-1644779 and the State of Florida.

# A  Pressure Control

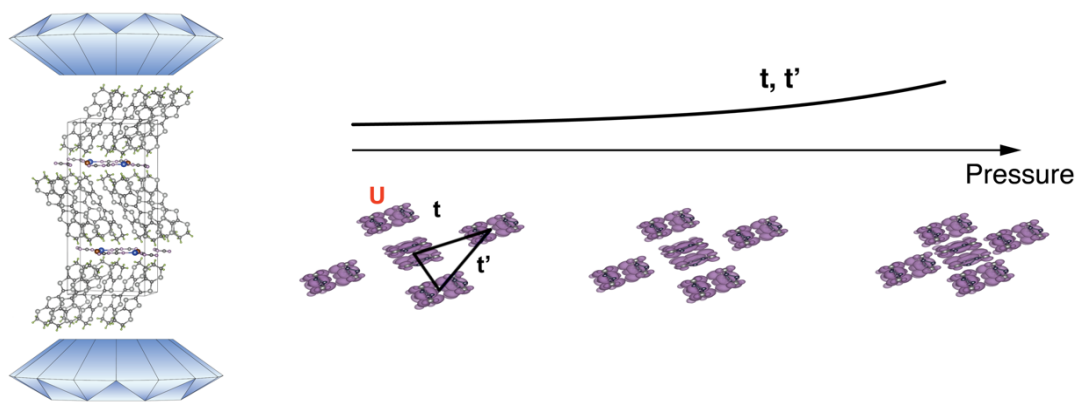

# B  Photo-molecular Control

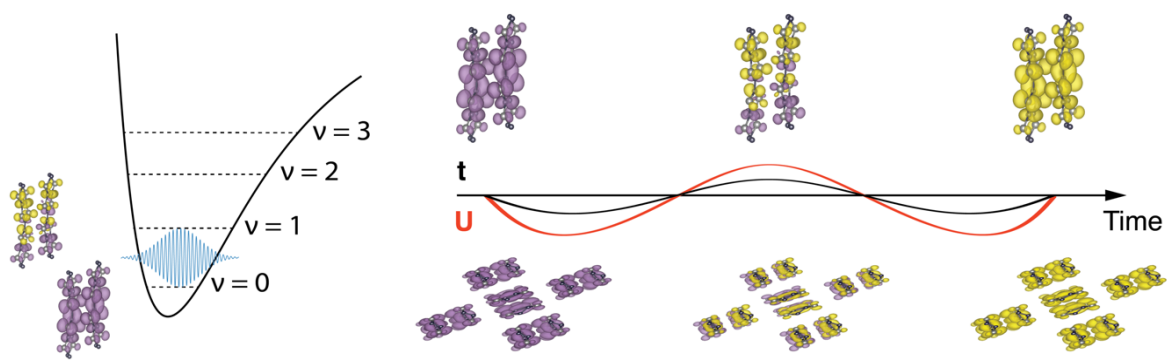

**Figure 1**: **a.** (left) Crystal structure of κ-(BEDT-TTF)$_2$Cu[N(CN)$_2$]Br. Alternating blocks of conducting and insulating layers are formed out of donor ET and polymeric chain-forming acceptor Cu[N(CN)$_2$]Br molecules, respectively. The ET molecules are paired in dimers. (right) View of the pressure-dependent arrangement of the dimers in the κ-phase (visualized along the long axis of the molecules). As (external or chemical) pressure is applied, the spacing between dimers changes, thus acting on the transfer integrals ($t$, $t'$) and tuning the ground state of the material. **b.** (left) Lennard-Jones potential alongside with highest occupied molecular orbital (HOMO) plots for the first two vibrational levels. (right) Time evolution of the HOMO after pulsed excitation creating a mixed population of $v = 1$ and $v = 0$ states, and corresponding arrangement of the ET dimers. The periodic changes in the wavefunction introduce modulations in the on-dimer Hubbard $U$ and transfer integral $t$, opening new dynamic pathways for controlling the material ground state.

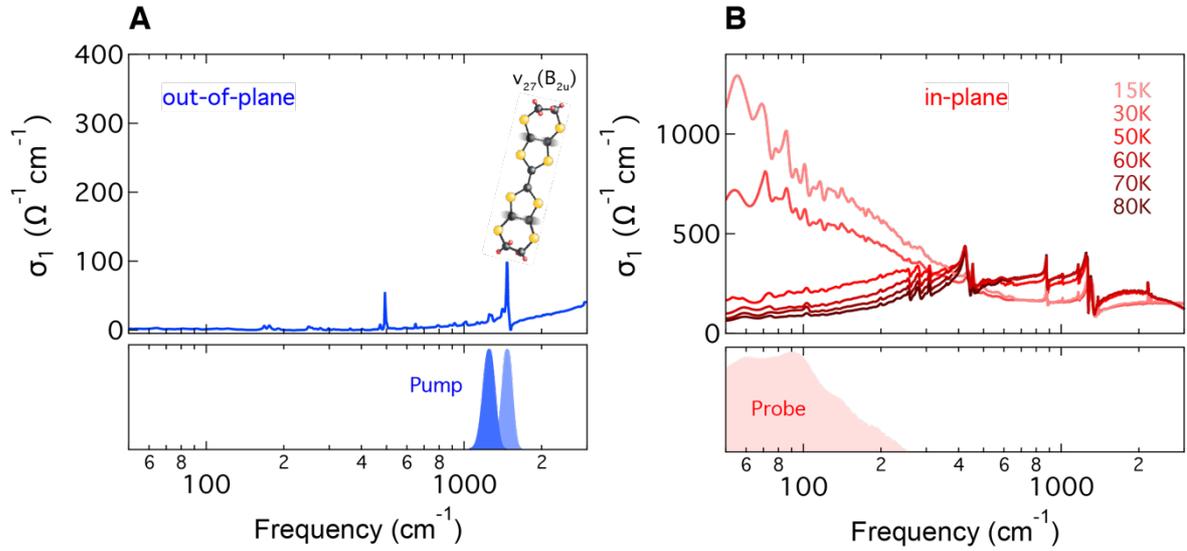

**Figure 2.** **a.** Upper panel: Out-of-plane equilibrium optical conductivity of κ-(ET)$_2$Cu[N(CN)$_2$]Br measured at $T$ = 15 K (*13*). Lower panel: Mid-infrared pump spectra tuned close to resonance with the ν$_{27}$ molecular vibration. The atomic motions corresponding to this mode are displayed in the upper panel (*16*). **b.** Upper panel: In-plane equilibrium optical conductivity measured at different temperatures between 15 K and 80 K. Lower panel: Frequency spectrum of the THz probe pulses used in our experiment.

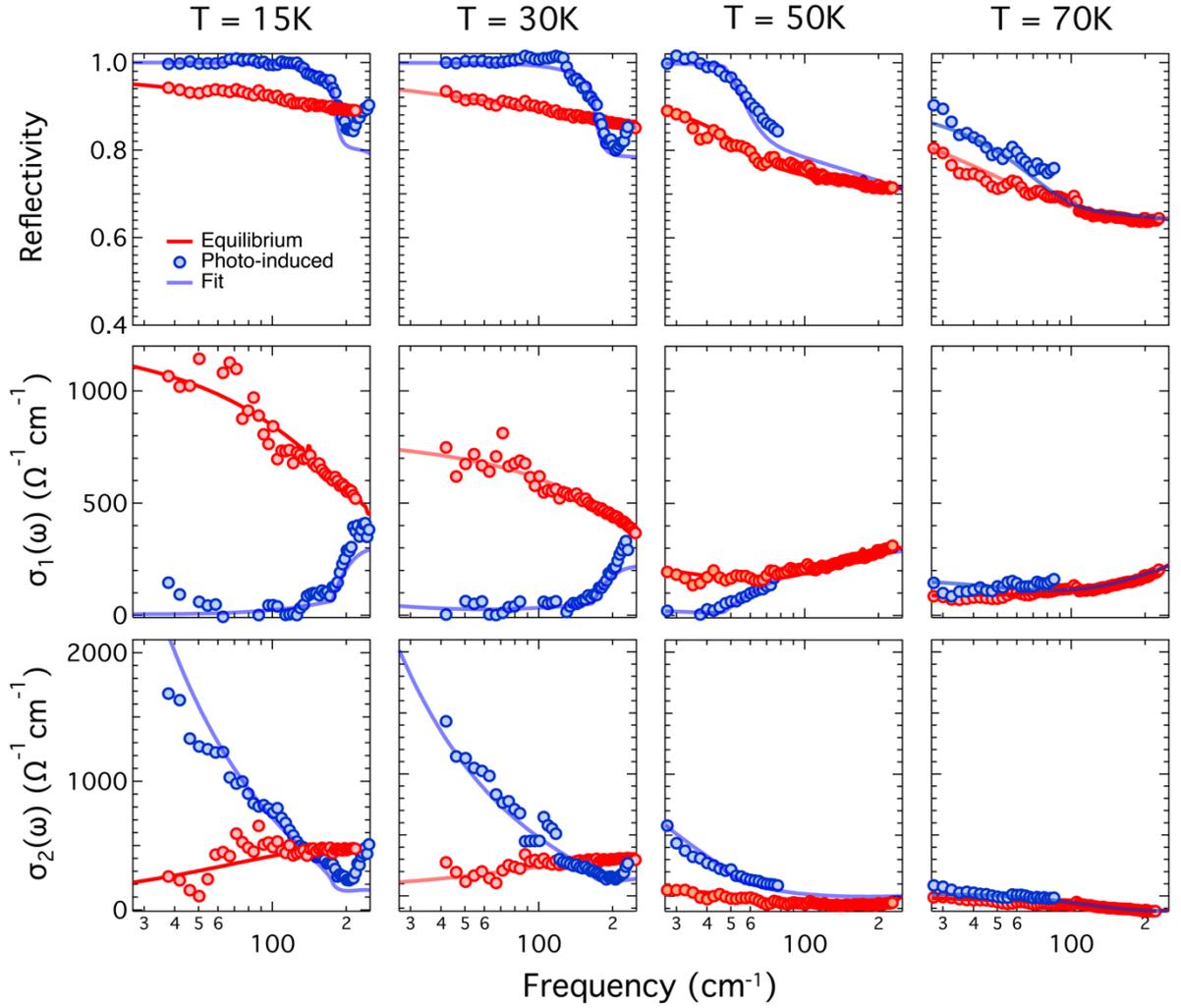

**Figure 3.** In-plane reflectivity, real and imaginary part of the optical conductivity measured in κ-(ET)$_2$Cu[N(CN)$_2$]Br at equilibrium (red circles) and at $\tau \simeq 1$ ps time delay after vibrational excitation (blue circles), at four different base temperatures between 15 K and 70 K. Lines are corresponding fits to the transient spectra. These were performed with a Drude-Lorentz model for the equilibrium response at all temperatures and for the transient spectra at T = 70 K. A Mattis-Bardeen model for superconductors was used instead for the out-of-equilibrium response at T ≲ 50 K. All data have been taken upon vibrational excitation close to resonance with the C=C stretching mode with a pump fluence of ∼2 mJ/cm$^2$.

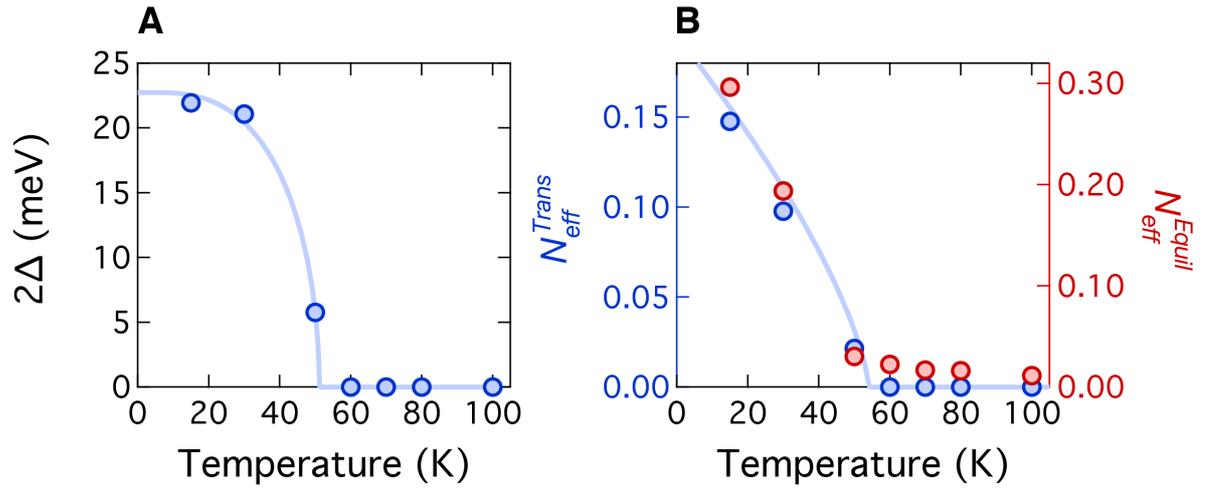

**Figure 4. a.** Photo-induced optical gap, 2Δ, extracted from Mattis-Bardeen fits to the transient optical properties of Fig. 3, as a function of base temperature. **b.** Blue, left scale: Temperature dependence of the effective number of "condensed" carriers per unit cell in the transient state, $N_{eff}^{Trans}$. Red, right scale: Corresponding effective number of mobile carriers per unit cell in the equilibrium metallic state before photo-excitation, $N_{eff}^{Equil}$ (see discussion in the text). These data have been taken upon vibrational excitation with a pump fluence of ~2 mJ/cm$^2$.

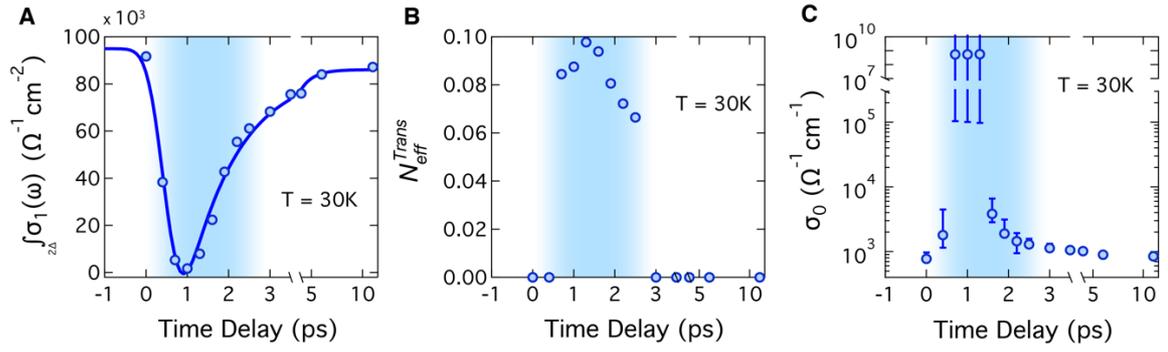

**Figure 5. a.** Dynamical evolution of the transient spectral weight in the real part of the optical conductivity as a function of pump-probe time delay. **b**. Time evolution of the effective number of "condensed" carriers per unit cell in the transient state, $N_{eff}^{Trans}$. **c.** Time evolution of the zero-frequency extrapolation of the optical conductivity, $\sigma_0 = \sigma_1(\omega)|_{\omega \to 0}$, extracted from Drude-Lorentz fits of the transient optical properties. Data have been taken at $T = 30$ K after vibrational excitation with a pump fluence of ~2 mJ/cm².

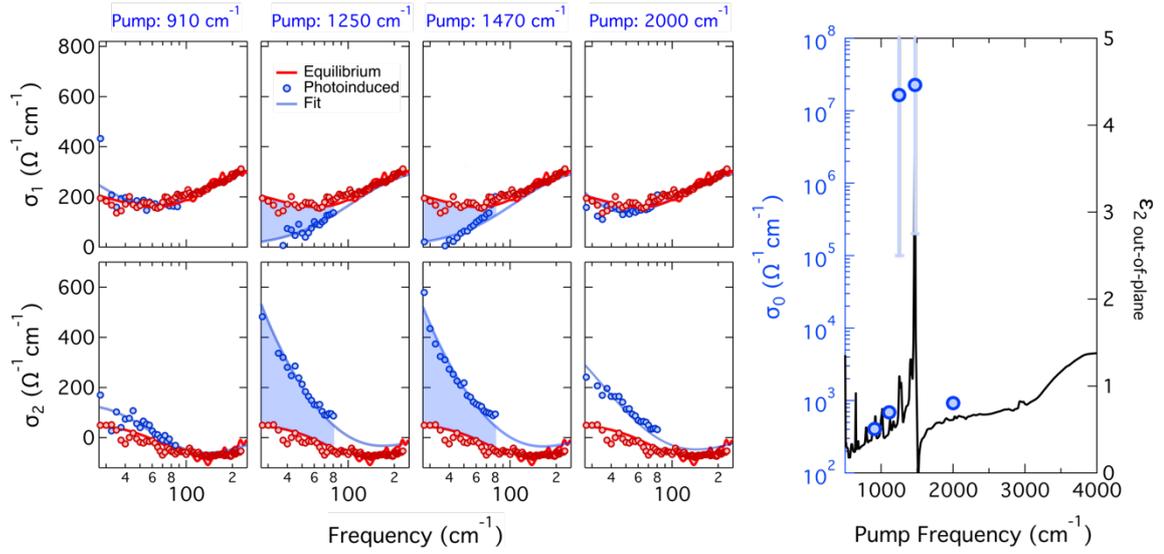

**Figure 6.** Left panels: In-plane complex optical conductivity of κ-(ET)$_2$Cu[N(CN)$_2$]Br measured at $T = 50$ K in equilibrium (red circles) and at $\tau \simeq 1$ ps time delay after photo-excitation (blue circles) at four different pump frequencies: $\nu_{pump} = 910$ cm$^{-1}$ ($\lambda_{pump} = 11$ μm), $\nu_{pump} = 1250$ cm$^{-1}$ ($\lambda_{pump} = 8$ μm), $\nu_{pump} = 1470$ cm$^{-1}$ ($\lambda_{pump} = 6.8$ μm), and $\nu_{pump} = 2000$ cm$^{-1}$ ($\lambda_{pump} = 5$ μm). Lines are corresponding fits performed with a Drude-Lorentz model. All data have been taken with a constant pump fluence of ~2 mJ/cm². Right panel: Zero-frequency extrapolation of the transient optical conductivity, $\sigma_0 = \sigma_1(\omega)|_{\omega \to 0}$ (see main text), extracted from the spectra on the left for different pump wavelengths (blue circles, left scale). The equilibrium out-of-plane imaginary dielectric function (*13*) is also displayed (black line, right scale) to visualize the resonant photo-conductivity behavior on the C=C ET vibration.

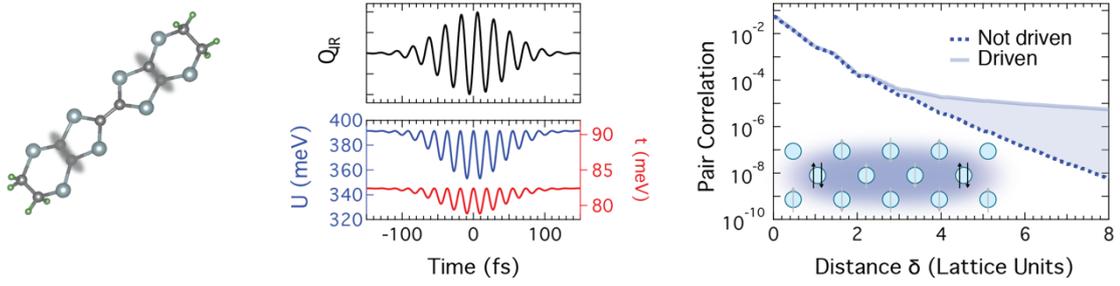

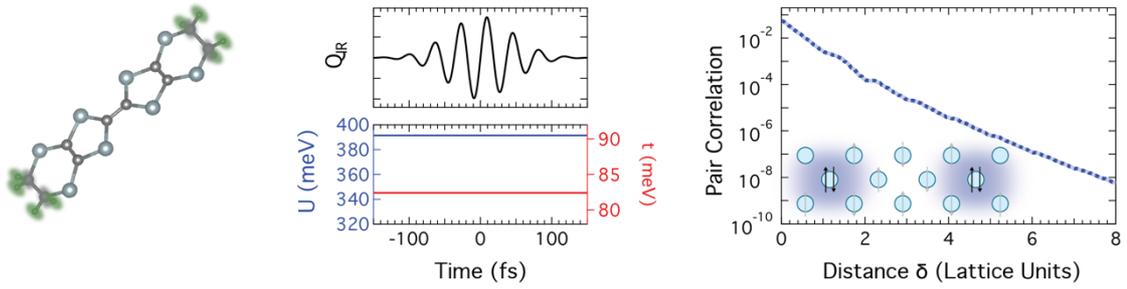

**Figure 7. (left panels)** Atomic motions corresponding to the 1470 cm$^{-1}$ (a) and 920 cm$^{-1}$ (b) modes of the ET molecule. **(center panels)** Time dependent displacement along the normal coordinates and induced modulations of the on-site Hubbard $U$ interaction and hopping matrix element $t$, calculated with ab initio simulations in the frozen-phonon approximation for the two different driven modes. **(right panels)** Corresponding spatial evolution of the pair correlation function, calculated with a time-dependent triangular ladder Fermi-Hubbard model in the undriven case (blue dashed line) and driven case (blue solid lines) for the two modes. The insets display a schematic representation of the correlation between pairs on different sites.

# Photo-molecular high temperature superconductivity

M. Buzzi[1,*], D. Nicoletti[1], M. Fechner[1], N. Tancogne-Dejean[1], M. A. Sentef[1],

A. Georges[2,3], M. Dressel[4], A. Henderson[5], T. Siegrist[5], J. A. Schlueter[5,6], K. Miyagawa[7],

K. Kanoda[7], M.-S. Nam[8], A. Ardavan[8], J. Coulthard[8], J. Tindall[8], F. Schlawin[8], D. Jaksch[8],

A. Cavalleri[1,8]

[1] *Max Planck Institute for the Structure and Dynamics of Matter, 22761 Hamburg, Germany*
[2] *Center for Computational Quantum Physics (CCQ), The Flatiron Institute, New York, NY, USA*
[3] *Collège de France, 11 place Marcelin Berthelot, 75005 Paris, France*
[4] *1. Physikalisches Institut, Universität Stuttgart, 70569 Stuttgart, Germany*
[5] *National High Magnetic Field Laboratory, 1800 E Paul Dirac Drive, Tallahassee, FL 31310, USA*
[6] *Division of Material Research, National Science Foundation, Alexandria, VA 22314, USA*
[7] *Department of Applied Physics, University of Tokyo, 7-3-1 Hongo, Bunkyo-ku, Tokyo 113-8656, Japan*
[8] *Department of Physics, Clarendon Laboratory, University of Oxford, Oxford OX1 3PU, United Kingdom*
* e-mail: michele.buzzi@mpsd.mpg.de

# Supplementary Material

**S1. Sample preparation and equilibrium optical properties**

**S2. Determination of the transient optical properties**

**S3. Fitting models**

**S4. Extended data sets**

**S5. Pump fluence dependence**

**S6. Calculation of the effective Hubbard parameters**

**S7. Driven Hubbard model**

# S1. Sample preparation and equilibrium optical properties

Single crystals of κ-(BEDT-TTF)$_2$Cu[N(CN)$_2$]Br with typical dimensions of 0.5 × 0.3 × 0.5 mm$^3$ were synthesized by electro-crystallization, following the procedure described in Refs. 1 & 2. The (*ac*)-axes define the highly conducting plane, while the *b*-axis is normal to that plane. The equilibrium superconducting transition at $T_C \simeq 12$ K was characterized with resistivity measurements(*1, 2*).

The terminal ethylene groups of the BEDT-TTF molecule (henceforth abbreviated ET) in this compound undergo an ordering transition, at about 80 K. To ensure minimum disorder of the ethylene groups, in all of our experiments the samples were cooled down at the same low rate of ∼150 mK/min below 160 K (*1, 2*).

The κ-(ET)$_2$Cu[N(CN)$_2$]Br equilibrium optical properties were measured using a Bruker Vertex 80v interferometer. The sample was mounted on the tip of a cone-shaped holder, with the in-plane crystal surface exposed to the beam. The holder was installed on the cold finger of a He-flow cryostat, thus enabling to collect broadband infrared spectra in reflection geometry at different temperatures. These spectra were then referenced against a thin gold film evaporated in-situ on the same sample surface.

The reflectivity curves obtained with this procedure, covering a range between ∼25 – 5000 cm$^{-1}$, were extrapolated to $\omega \to 0$ with Drude fits (see also Section S3), and extended to higher frequencies with literature data (*3, 4*). This allowed us to perform Kramers-Kronig transformations and retrieve full sets of in-plane equilibrium response functions to be used as reference in our pump-probe experiment.

Examples of selected reflectivity and complex optical conductivity spectra are reported in Fig. S1. As discussed in the main text, the normal state equilibrium response is characterized by a metallic Drude peak (see Fig. S1b) for temperatures below $T^* \simeq 50$ K, while the spectra at higher temperatures are those of a "bad metal", characterized by a non-zero low-frequency conductivity, in absence of a Drude peak (see also Ref. 3).

We did not observe any change in the optical spectra across the superconducting transition temperature $T_C = 12.5$ K. The absence of a clear optical gap in the superconducting κ-(ET)$_2$X compounds has already been reported in the past (*5-8*).

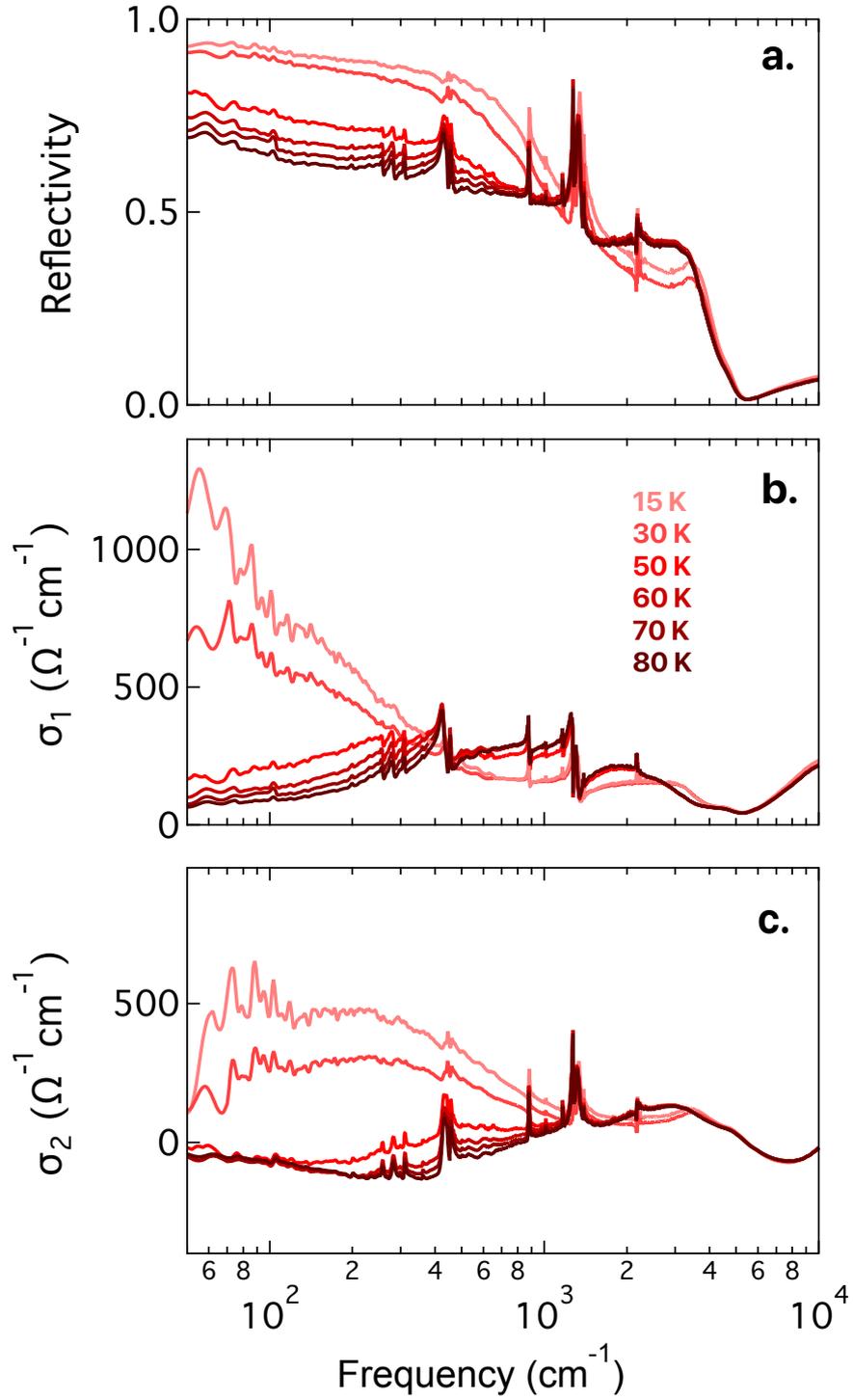

**Fig. S1.** Equilibrium in-plane reflectivity (a), real (b), and imaginary part (c) of the optical conductivity of κ-(ET)$_2$Cu[N(CN)$_2$]Br, measured at different temperatures following the procedure describe in the text.

## S2. Determination of the transient optical properties

In a series of mid-infrared pump / THz probe experiments, we investigated different κ-(ET)$_2$Cu[N(CN)$_2$]Br single crystals. Each of them was glued on the top of a cone-shaped sample holder, exposing a crystal face which contained both an in-plane and an out-of-plane axis (with ~0.5 mm and ~0.3 mm dimension, respectively). The holder was installed on the cold finger of a He-flow cryostat for temperature dependent measurements.

κ-(ET)$_2$Cu[N(CN)$_2$]Br was photo-excited with ~100 fs long mid-infrared pulses, tuned to different wavelengths ($\lambda_{pump} \simeq$ 11 μm, 8 μm, 6.8 μm, 5 μm), some of which were resonant with local vibrational modes of the ET molecules. These pump pulses were generated by difference frequency mixing of the signal and idler outputs of an optical parametric amplifier (OPA) in a 0.5 mm thick GaSe crystal. The OPA was pumped with ~100 fs long pulses from a commercial Ti:Sapphire regenerative amplifier (800-nm wavelength).

The pump pulses were focused onto the sample surface, with their polarization aligned along the out-of-plane axis. The typical pump spot size was ~0.5 mm, allowing for a full illumination of the crystal surface. A maximum fluence of ~4 mJ/cm$^2$ could be achieved, corresponding to a peak electric field of ~4 MV/cm.

The transient reflectivity changes after photo-excitation were determined via time-domain THz spectroscopy in two different experimental setups. Single-cycle THz pulses were generated either in 1-mm thick ZnTe or in 0.2-mm thick GaP, using 100-fs and 30-fs long 800 nm pulses, respectively. These probe pulses were focused at normal incidence onto the sample surface with polarization along the in-plane direction, and their electric field profile was measured, after reflection, via electro-optic sampling in nonlinear crystals identical to those used for THz generation (*i.e.*, 1-mm thick ZnTe and 0.2-mm thick GaP). The ZnTe based setup produced a spectral bandwidth extending from ~25 to 70 cm$^{-1}$, while the GaP based setup covered the range from ~40 to 230 cm$^{-1}$.

In order to minimize the effects on the pump-probe time resolution due to the finite duration of the THz probe pulse, we performed the experiment as described in Ref. 9. The transient reflected field at each time delay $\tau$ after excitation was obtained by keeping fixed the delay $\tau$ between the pump pulse and the electro-optic sampling gate pulse, while scanning the delay $t$ of the single-cycle THz probe pulse.

The stationary probe electric field $E_R(t)$ and the differential electric field $\Delta E_R(t,\tau)$ reflected from the sample were recorded simultaneously by feeding the electro-optic sampling signal into two lock-in amplifiers and mechanically chopping the pump and probe beams at different frequencies. $E_R(t)$ and $\Delta E_R(t,\tau)$ were then independently Fourier transformed to obtain the complex-valued, frequency-dependent $\tilde{E}_R(\omega)$ and $\Delta\tilde{E}_R(\omega,\tau)$. The photo-excited complex reflection coefficient $\tilde{r}(\omega,\tau)$ was determined by

$$\frac{\Delta\tilde{E}_R(\omega,\tau)}{\tilde{E}_R(\omega)} = \frac{\tilde{r}(\omega,\tau) - \tilde{r}_0(\omega)}{\tilde{r}_0(\omega)},$$

where $\tilde{r}_0(\omega)$ is the stationary reflection coefficient known from the equilibrium optical response (see Section S1).

As the penetration depth of the excitation pulses was typically larger than that of the THz probe pulses (with the only exception of the 6.8 µm pump experiment, see Figure S2), the THz probe pulse sampled a homogeneously excited volume. The transient optical properties could then be extracted directly, without the need to consider any pump-probe penetration depth mismatch in the data analysis.

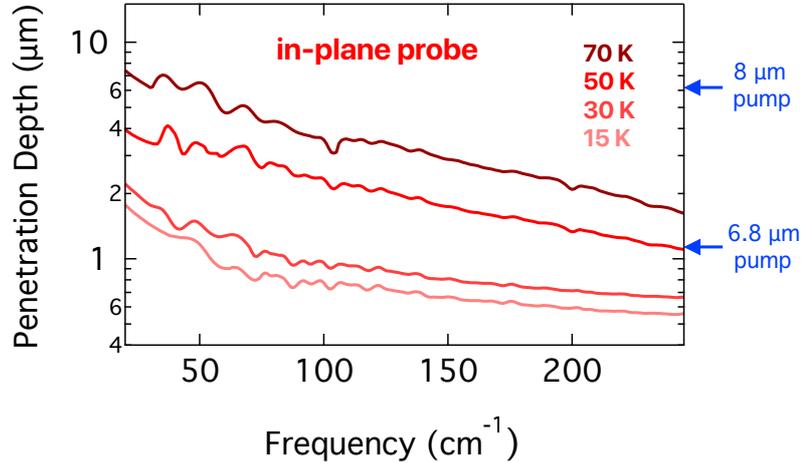

**Fig. S2.** Field penetration depth, $d(\omega) = \frac{c}{\omega \cdot \mathrm{Im}[\tilde{n}_0(\omega)]}$ (here $\tilde{n}_0(\omega)$ is the stationary complex refractive index), calculated from the in-plane equilibrium optical properties of κ-(ET)$_2$Cu[N(CN)$_2$]Br (see Fig. S1), at selected temperatures. Data are displayed over the frequency range covered by our THz probe pulses. Arrows on the right axis indicate the corresponding out-of-plane pump penetration depth values for two selected excitation frequencies, extracted from the data in Ref. 10.

In this case, the complex refractive index of the photo-excited material, $\tilde{n}(\omega,\tau)$, was directly retrieved from the Fresnel relation:

$$\tilde{n}(\omega,\tau) = \frac{\tilde{r}(\omega,\tau) - 1}{\tilde{r}(\omega,\tau) + 1},$$

and from this, the transient complex optical conductivity, $\tilde{\sigma}(\omega,\tau) = \frac{\omega}{4\pi i}[\tilde{n}(\omega,\tau)^2 - \varepsilon_\infty]$.

For the limited set of data that required a pump-probe penetration depth analysis ($\lambda_{pump} = 6.8\ \mu m, T \geq 50\ K$), we followed the procedure described in Refs. 11 & 12. We treated the photo-excited surface as a stack of thin layers with a homogeneous refractive index and described the excitation profile by an exponential decay. By numerically solving the coupled Fresnel equations of such multi-layer system, the refractive index at the surface could be retrieved, and from this the complex conductivity for a homogeneously transformed volume.

Importantly, this renormalization only affected the size of the response, whereas the qualitative changes in optical properties were independent of it and the specific model chosen (*12*).

## S3. Fitting models

The in-plane equilibrium optical properties of κ-(ET)$_2$Cu[N(CN)$_2$]Br were fitted at all measured $T > T_C$ with a Drude-Lorentz model, for which the complex optical conductivity is expressed as:

$$\tilde{\sigma}_{DL}(\omega) = \frac{\omega_p^2}{4\pi} \frac{1}{\gamma_D - i\omega} + \sum_i \frac{\Omega_{p,i}^2}{4\pi} \frac{\omega}{i(\Omega_{0,i}^2 - \omega^2) + \Gamma_i \omega}. \qquad (S3.1)$$

Here, $\omega_p$ and $\gamma_D$ are the Drude plasma frequency and momentum relaxation rate, while $\Omega_{0,i}, \Omega_{p,i},$ and $\Gamma_i$ are the peak frequency, plasma frequency, and damping coefficient of the *i*-th oscillator, respectively.

The same Drude-Lorentz model was also employed to fit the transient optical spectra after photo-excitation. Here, all parameters related to the oscillators at frequencies outside the measurement range were kept fixed to the values determined at equilibrium, while the lowest frequency oscillator and the Drude term were left free to vary. Importantly, for each data set, the same parameters were used to simultaneously fit the reflectivity, the real part of the optical conductivity, as well as its imaginary part. As shown, for example, in Fig. 3 & Fig. 6 of the main text, this Drude-Lorentz model is able to fully reproduce the experimental data for the temperatures and time delays that show a metallic behavior and no signature of a superconducting-like response.

In addition, the transient superconducting-like response of κ-(ET)$_2$Cu[N(CN)$_2$]Br could also be captured (see fits in Fig. 6 of main text) by the simple $\gamma_D \to 0$ limit of Eq. S3.1,

$$\tilde{\sigma}_{DL}(\omega, \gamma_D \to 0) = \frac{\pi}{2}\frac{N_S e^2}{m}\delta[\omega = 0] + i\frac{N_S e^2}{m}\frac{1}{\omega} + \sum_i \frac{\Omega_{p,i}^2}{4\pi}\frac{\omega}{i(\Omega_{0,i}^2 - \omega^2) + \Gamma_i \omega}. \quad (S3.2)$$

Here $N_S$, $e$, and $m$ are the superfluid density, electron charge, and electron mass, respectively.

That said, the Drude-Lorentz model of Eq. S3.1 could be applied to all transient optical spectra reported in this work, without any assumption on the nature of the non-equilibrium state (see also Ref. 13). The zero-frequency limit $\sigma_0 = \lim_{\omega \to 0} \tilde{\sigma}_{DL}(\omega)$ extracted from these fits (see Fig. 5 & Fig. 6 of main text), which is a finite quantity in a metal and diverges in a "perfect conductor", was used as a discriminator for the presence of transient superconductivity (*13*).

Additional fitting of the superconducting-like optical properties was performed with an extension of the Mattis-Bardeen model for superconductors of variable purity (*14, 15*), which is typically used to describe the response of superconductors at equilibrium, for finite frequencies and temperatures. Fit curves extracted with this procedure are reported, for example, in Fig. 3 of the main text. The corresponding values of the optical gap, 2Δ(*T*), extracted with this procedure are shown instead in Fig. 4a.

## S4. Extended data sets

We report here a comparison of transient optical spectra taken under the same conditions (temperature, excitation wavelength and fluence, pump-probe time delay) on three different κ-(ET)$_2$Cu[N(CN)$_2$]Br crystal coming from the same batch of samples. These data, measured at $T = 30$ K, are reported in Fig. S4. Therein, we show how, aside from small differences, the non-equilibrium response appears to be sample independent, and all signatures of transient, photo-induced superconductivity, i.e., a reflectivity equal to 1, a gap in $\sigma_1(\omega)$, and a ~$1/\omega$ divergence in $\sigma_2(\omega)$, are fully preserved.

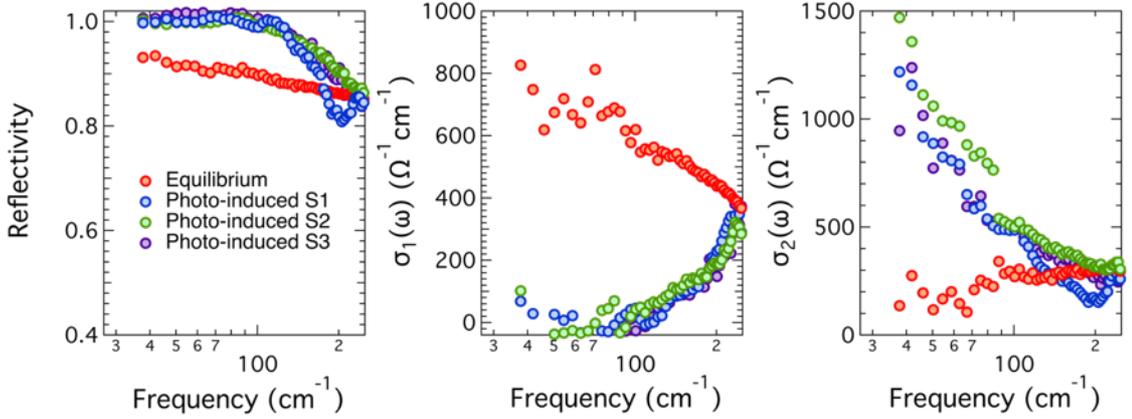

**Fig. S4.** Equilibrium in-plane reflectivity, real, and imaginary part of the optical conductivity, measured at $T = 30$ K under the very same excitation conditions ($\lambda_{pump} = 8$ μm, ~2 mJ/cm² fluence) on three different κ-(ET)$_2$Cu[N(CN)$_2$]Br crystals, at $\tau = 1$ ps pump-probe time delay (blue, green, and purple circles). Red data points are the spectra at equilibrium.

## S5. Pump fluence dependence

In Figure S5 we show pump fluence dependent data. In analogy with Fig. 4 of the main text, we plot the non-equilibrium "superfluid density", determined from the low-frequency extrapolation of the transient imaginary conductivity as $N_{eff}^{Trans} = \frac{mV_{Cell}}{e^2} \lim_{\omega \to 0}[\omega\sigma_2^{Trans}(\omega)]$, where $m$ is the bare electron mass, $V_{Cell}$ the unit cell volume, and $e$ the electron charge. This effective number of "condensed" carriers displays a monotonic increase with increasing excitation strength, with signatures of a saturation for fluences above ~2 mJ/cm².

The saturation seems to occur for a value of $N_{eff}^{Trans}$ that approaches the equilibrium quasi-particle density, $N_{eff}^{Equil} = \frac{mV_{Cell}}{4\pi e^2}\left(\omega_P^{Equil}\right)^2$, which we show for reference as a horizontal dashed line (here $\omega_P^{Equil}$ is the equilibrium carrier plasma frequency extracted from Drude-Lorentz fits).

This observation, combined with the temperature dependent data of Fig. 4b of the main text, is suggestive of an intimate connection between the photo-excited carriers in the transient superconducting state and pre-existing quasiparticles at equilibrium.

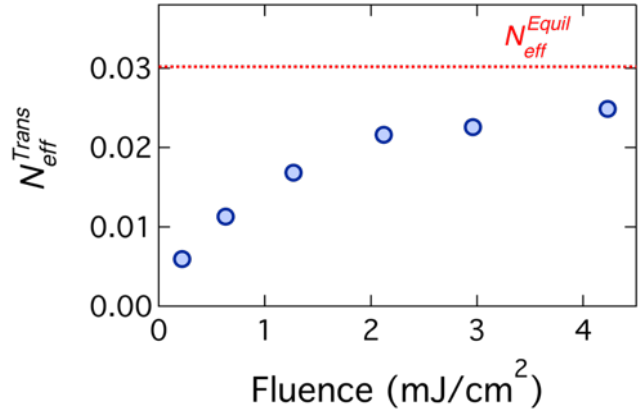

**Fig. S5.** Pump fluence dependence of the effective number of "condensed" carriers per unit cell in the transient state, $N_{eff}^{Trans}$ (blue circles), shown along with the effective number of mobile carriers in the equilibrium metallic state before photo-excitation, $N_{eff}^{Equil}$ (horizontal dashed line, see discussion in the text). These data have been taken upon 8 μm excitation at T = 50 K.

# S6. Calculation of the effective Hubbard parameters

**Geometry optimization**

We performed a geometry optimization of the crystal structure for κ-Br in order to obtain relaxed atomic coordinates. This was achieved using the density functional theory code Quantum ESPRESSO (*16, 17*) starting from the x-ray crystallographic data of Ref 18. We used the PBE functional complemented by the Grimm D2 van der Waals correction, norm-conserving pseudopotentials, a $2 \times 1 \times 3$ momentum grid, an energy cutoff of 1224 eV, and a finite-temperature smearing of 10 meV. Forces were optimized to be smaller than $10^{-4}$ Ha/Bohr. We obtained relaxed lattice constants of 12.816 Å, 29.608 Å, and 8.493 Å, for the *a*, *b*, and *c* crystallographic directions, respectively. These values are in excellent agreement with reported measured values (*18, 19*).

**Phonon mode calculations**

We computed the phonon modes of the isolated ET dimer and constructed the $B_u$ modes by applying the corresponding symmetries (*20*). These $B_u$ modes are the infrared-active modes with dipole along the out-of-plane *b* direction. We obtained the eigenvectors of the dynamical matrix from Quantum ESPRESSO, using the same parameters and cell as for the geometry optimization of the full κ-Br crystal.

Figure S6.1 shows examples of eigenvectors for two modes with frequencies close to the experimental excitation conditions. While the 10.7 μm vibration (terminal ethylene mode) involves motions of the ethylene groups at the ends of each ET molecule, the 6.8 μm mode (C=C stretching mode) acts almost exclusively on the C atoms.

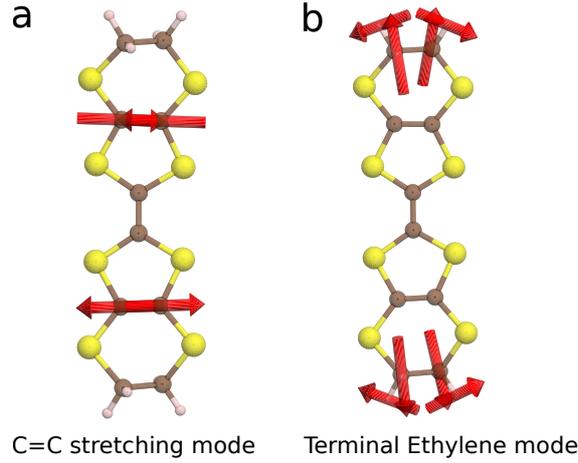

C=C stretching mode     Terminal Ethylene mode

**Fig. S6.1.** Calculated $B_u$ phonon mode eigenvectors corresponding to excitation wavelengths considered in this work. Here, the modes are displayed only for a single ET molecule.

**Extraction of ground-state Hubbard model parameters**

The effective electronic parameters of the weak-bond triangular lattice Hubbard model were obtained following the approach proposed in Ref. 21. To this end we computed the ground state and corresponding band structure using the Octopus code (*22, 23*). We employed the same norm-conserving pseudopotentials, PBE functional, and k-point grid as for the geometry optimization. The real-space grid was sampled with a spacing of 0.3 Bohr. The resulting band structure (Fig. S6.2) was fitted by a tight-binding model for 8 ET sites, leading to four parameters ($t_1 - t_4$), and fixing the electronic occupation to 3/4 filling (*24*). From these fits, we extracted the effective parameters $U$, $t$, and $t'$ of the single band Hubbard model (*21, 24*) involving only the electronic band that crosses the Fermi level in Fig. S6.2.

$$H = \sum_{<ij>,\sigma} t\left(c_{i\sigma}^\dagger c_{j\sigma} + H.c.\right) + \sum_{[ij],\sigma} t'\left(c_{i\sigma}^\dagger c_{j\sigma} + H.c.\right) + U \sum_i \left(n_{i\uparrow} - \frac{1}{2}\right)\left(n_{i\downarrow} - \frac{1}{2}\right)$$

Note that the single-band Hubbard model is half-filled for the ¾-filled two-band tight-binding model.

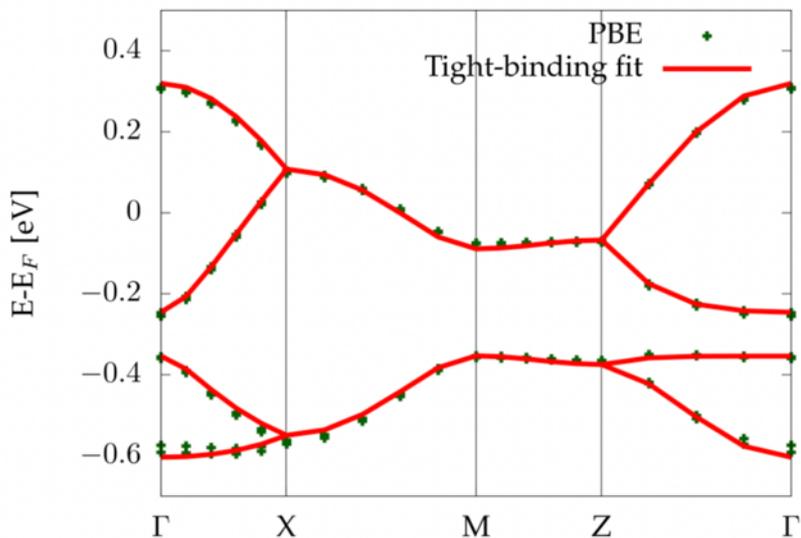

**Fig. S6.2.** Ground-state band structure of κ-Br calculated at the PBE level and tight-binding fit.

For κ-Br, the Cu 3d bands overlap with the ET molecule bands, complicating the extraction of the tight-binding parameters, as noted in Ref. 21, 25. As DFT with semi-local functionals suffers from the so-called delocalization problem, and 3d transition metal bands are not well-described, we included a Hubbard $U$ of 7 eV on the Cu 3d orbitals, which has the effect of improving the description of the electronic structure and it also shifts down the Cu 3d bands, thus making the fitting procedure more robust. For the equilibrium, undistorted structure, we obtained values of $U/t = 4.73$ and $t'/t = 0.24$. The $U/t$ value is comparable to that reported in Ref. 21, but $t'/t$ comes out smaller. This deviation is attributed to our inclusion of the van der Waals corrections for the geometry relaxation, which were not considered in Ref. 21. The van der Waals corrections significantly affect the volume of the relaxed crystal.

**Extraction of Hubbard model parameters for displaced structures**

Starting from the geometrically relaxed structure, we systematically displaced the ions according to the phonon mode coordinates obtained in the calculations. For each displaced structure, we computed the adiabatic electronic ground state and band structure in the frozen-phonon approximation using the Octopus code with the same technical parameters employed in the ground state calculation. We then followed exactly the same steps as in the ground state case to perform the tight-binding fits and finally extracted the effective Hubbard model parameters. Figure S6.3 shows the resulting modulations of $U$, $t$, and the ratios $U/t$ and $t'/t$ for displacements along the $B_u$ mode coordinates.

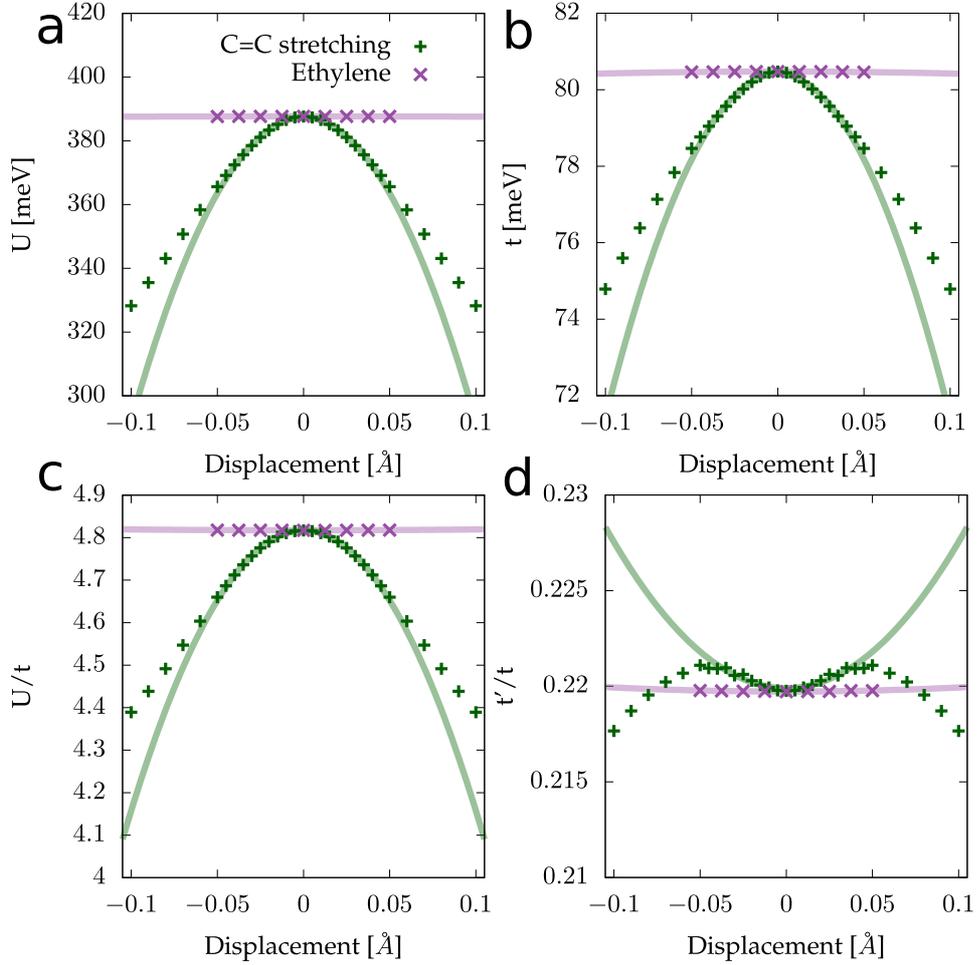

**Fig. S6.3.** Hubbard model parameters as a function of mode displacement. **(a)** On-dimer Hubbard interaction $U$, **(b)** strong bond hopping integral, $t$, **(c)** ratio $U/t$, and **(d)** t'/t. In each plot the lines represent parabolic fits to the data. Structural deformations along the normal coordinates of the C=C stretch induce strong, non-linear modifications of the electronic interaction parameters (green lines and symbols). By contrast, displacement of the terminal ethylene mode does not induce significant changes (purple lines and symbols).

## S7. Driven Hubbard model

The geometry of the triangular Fermi-Hubbard model considered here is shown in Fig. S7.1. The driven Hamiltonian is given by:

$$H(\tau) = \sum_{<ij>,\sigma} t(\tau)\left(c_{i\sigma}^\dagger c_{j\sigma} + H.c.\right)$$
$$+ \sum_{[ij],\sigma} t'\left(c_{i\sigma}^\dagger c_{j\sigma} + H.c.\right) + U(\tau)\sum_i \left(n_{i\uparrow} - \frac{1}{2}\right)\left(n_{i\downarrow} - \frac{1}{2}\right), \quad (S7.1)$$

where $\tau$ denotes the time, $t$ is the nearest neighbour hopping element and $t'$ is the hopping element in the vertical direction. Following the frozen phonon simulations reported in Section S6, we assumed the on-site interaction $U$ and the strong bond hopping elements $t$ to be modulated by the phonon driving as follows:

$$U(\tau) = U_0 * \left(1 - A_1 \sin^2(\Omega\tau) e^{-\frac{(\tau-\tau_0)^2}{(2\tau_w)^2}}\right),$$

$$t(\tau) = t_0 * \left(1 - A_2 \sin^2(\Omega\tau) e^{-\frac{(\tau-\tau_0)^2}{(2\tau_w)^2}}\right). \quad (S7.2)$$

Here, the amplitudes $A_1$ and $A_2$ were extracted from the calculations in Section S6, the modulation frequency $\Omega$ was set to the phonon frequency, while the parameters $\tau_w$ and $\tau_0$ control the duration and delay of the sinusoidal pulse. We assumed that the vertical hopping strength $t'$ remains constant.

In Fig. S7.2 we report surface plots of the doublon correlations in distance and time for various system sizes. For sufficiently large systems we observe the emergence of uniform long-range doublon correlations which stabilise and persist over the simulation timescale. The increase in long-range correlation strength becomes more pronounced as the system size increases.

Finally, in Fig. 3 we show the full time-dynamics of the doublon correlations for various driving strengths and a fixed system size.

The matrix product calculations used to produce these results were performed using the Tensor Network Library (*26*). The system was initialised in its ground state using the Density Matrix Renormalisation Group algorithm (*27*) whilst the time evolution was performed with the Time Evolving Block Decimation method (*28*) on the resulting initial state. We used a second order Suzuki-Trotter decomposition of the time evolution operator with a time-step of $\delta\tau t = \pi\Omega/50$. In all our calculations we ensured that our results remained unchanged upon increasing the bond dimension from the specified value in the figures.

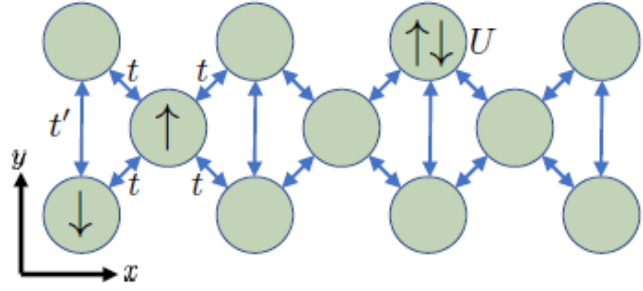

**Fig. S7.1.** A sketch of the geometry of the Hubbard ladder investigated in our study. Electrons can hop between nearest neighbour sites with amplitude $t$, and vertically with amplitude $t'$. Two electrons occupying the same site experience a repulsion $U$.

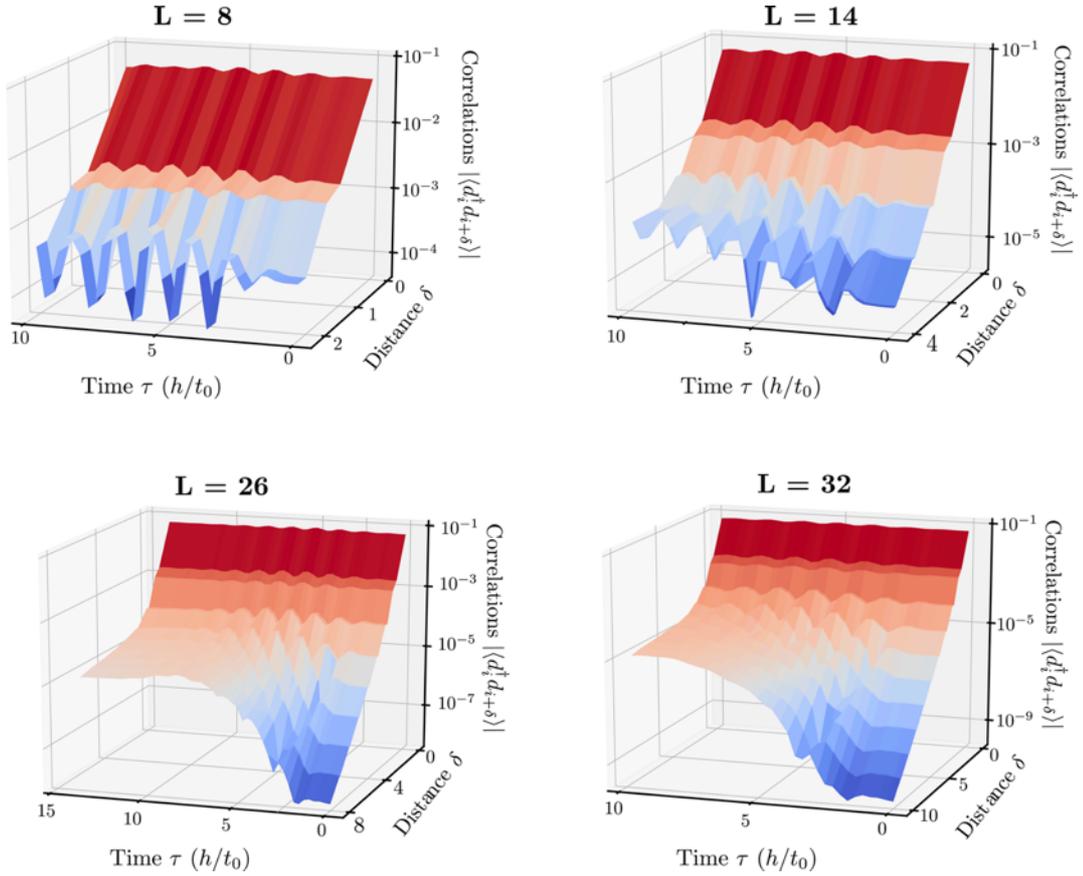

**Fig. S7.2.** Doublon Correlations as a function of distance (in lattice units) and time (in units of $h/t_0$) for the triangular ladder Hubbard model described by the Hamiltonian in Eq. (1). The four panels show calculations for four different system sizes $L$. The system is initialised in its ground state with $U_0 = 4.82\, t_0, t' = 0.22 t_0$ and half-filling. It was then allowed to evolve under time-modulated interaction strengths (see Eq. S7.2) with driving parameters $\tau_0 = 5.0 t_0$, $\tau_w = 5.0 t_0$, $\Omega = 2.27 t_0$, $A_1 = 0.155$ and $A_2 = 0.07$.

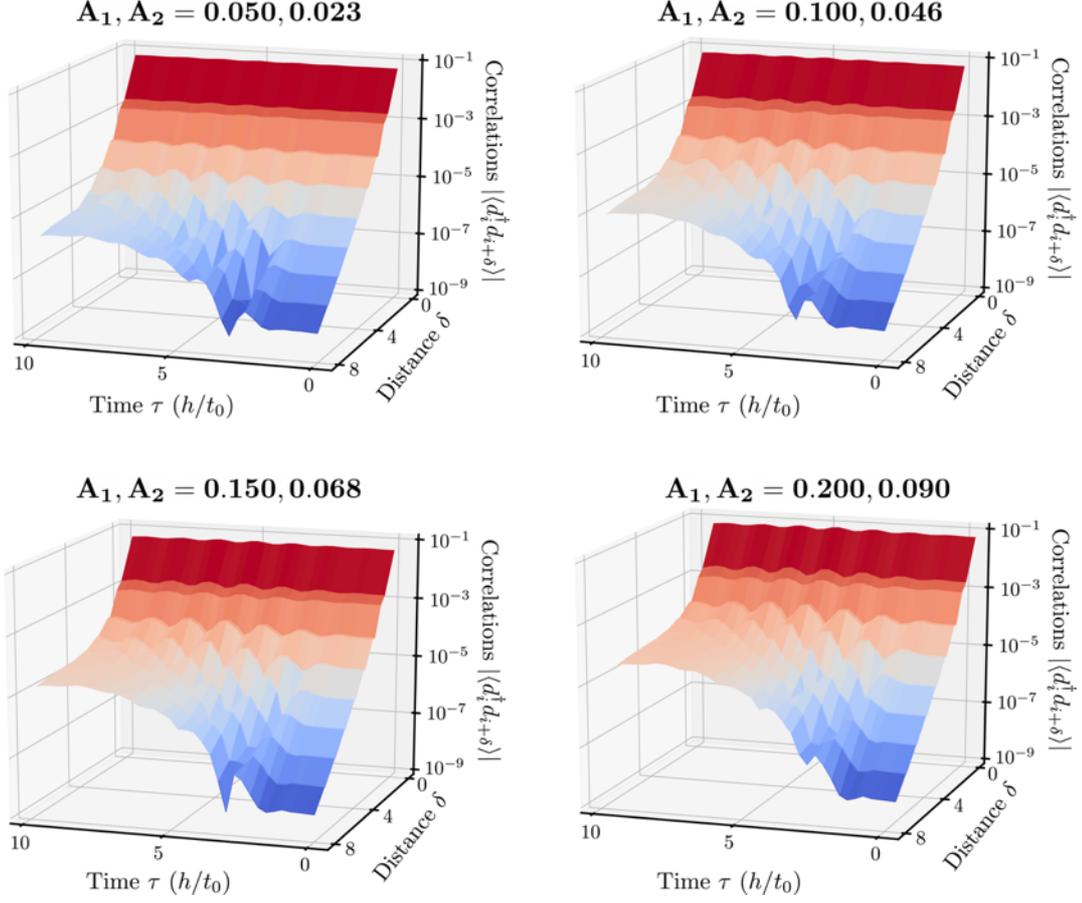

**Fig. S7.3.** Doublon correlations as a function of distance (in lattice units) and time (in units of $h/t_0$) for the 26-site triangular ladder Hubbard model described by the Hamiltonian in Eq. S7.1. The four panels show calculations for different driving amplitudes $A_1, A_2$. The system was initialised in its ground state with $U_0 = 4.73 t_0$, $t' = 0.24 t_0$, at half filling. It was then allowed to evolve under time-modulated interaction strengths (see Eq. S7.2) with driving parameters $\tau_0 = 5.0 t_0$, $\tau_w = 5.0 t_0$, $\Omega = 2.27 t_0$, and constant ratio $A_1/A_2 = 2.21$.

# References (Supplementary Material)